\documentclass[epsf,prx,twocolumn,showpacs,nofootinbib]{revtex4-1}

\usepackage[pdftex]{graphicx}
\usepackage{dcolumn}
\usepackage{bm}
\usepackage{epsfig}
\usepackage{latexsym}
\usepackage{amsmath}
\usepackage{amsfonts}
\usepackage{amssymb}
\usepackage{color}
\usepackage{array}
\usepackage{framed}

\begin{document}

\title{Localization of Extended Quantum Objects}
\author{Michael Pretko and Rahul M. Nandkishore \\
\emph{Department of Physics and Center for Theory of Quantum Matter, \\
University of Colorado, Boulder, CO 80309}}
\date{\today}

\begin{abstract}
A quantum system of particles can exist in a localized phase, exhibiting ergodicity breaking and maintaining forever a local memory of its initial conditions.  We generalize this concept to a system of extended objects, such as strings and membranes, arguing that such a system can also exhibit localization in the presence of sufficiently strong disorder (randomness) in the Hamiltonian.  We show that localization of large extended objects can be mapped to a lower-dimensional many-body localization problem.  For example, motion of a string involves propagation of point-like signals down its length to keep the different segments in causal contact.  For sufficiently strong disorder, all such internal modes will exhibit many-body localization, resulting in the localization of the entire string.  The eigenstates of the system can then be constructed perturbatively through a convergent `string locator expansion.'  We propose a type of out-of-time-order string correlator as a diagnostic of such a string localized phase.  Localization of other higher-dimensional objects, such as membranes, can also be studied through a hierarchical construction by mapping onto localization of lower-dimensional objects. Our arguments are `asymptotic' ($i.e.$ valid up to rare regions) but they extend the notion of localization (and localization protected order) to a host of settings where such ideas previously did not apply. These include high-dimensional ferromagnets with domain wall excitations, three-dimensional topological phases with loop-like excitations, and  three-dimensional type-II superconductors with flux line excitations. In type-II superconductors, localization of flux lines could stabilize superconductivity at energy densities where a normal state would arise in thermal equilibrium.
\end{abstract}

\maketitle

\tableofcontents

\section{Introduction}

Most textbook treatments of quantum statistical mechanics assume that the system under consideration has the ability to reach thermal equilibrium, mediated by coupling to an external heat bath.  Even for unitary evolution of a quantum system in isolation, the system itself can serve as a heat bath which thermalizes each of its own subsystems, in accordance with the Eigenstate Thermalization Hypothesis (ETH) \cite{srednicki,deutsch,rigol}.  From a modern perspective, however, we now know that it is possible for an isolated quantum system to fail to thermalize, in violation of the ETH, and to maintain a memory of its initial conditions at arbitrarily long times.  For example, in the presence of sufficiently strong disorder, a quantum system of particles can exhibit the phenomenon of many-body localization.

The physics of localization dates back to Anderson, who studied the single-body problem of non-interacting particles moving in a random potential\cite{anderson}.  In that seminal work, it was shown that strong disorder can transform the single-particle eigenstates from propagating plane waves into states which are localized at fixed positions in space, up to exponentially decaying tails.  The system is then unable to transport energy or other quantum numbers like charge, leading to vanishing dc conductivities.  While this single-particle picture is appealing, it is far from obvious that these results should carry over to an interacting many-body system.  Nevertheless, it has been shown that, under appropriate conditions, localization can indeed remain intact in the presence of interactions \cite{gornyi,basko1,basko2,imbrie,ogan,xxz}.  This discovery has led to a period of intense research on the subject of many-body localization (MBL), both in theory\cite{review,pal,moore,bauer,serbyn,swingle,order,growth,spt,huse,marginal,rahul,laumann,meanfield,tarun,khemani,protection,ros,gopal,bahri,vosk,finite,criterion,randomfield,universal,proximity,symmetry,deroeck,spectral,zhang,critical,chandran,bath,spectrum,power,italian,stability,highly,continuum,gaugembl,zhicheng,dumitrescu,class,geraedts,abhinav,spectrum2,acevedo,mobile,longrange}
and in experiments \cite{exp1,exp2,exp3,exp4}.

To date, the study of localization has almost exclusively focused on systems with point-like quasiparticles.  But in principle, the concept of localization should be applicable to more general classes of objects.  In particular, one could consider systems with excitations taking the form of extended objects, such as loop excitations in a three-dimensional topological phase, or flux lines in a three-dimensional type-II superconductor.  The dynamics of such objects are fairly complicated to describe, even in a clean system, and it is not immediately clear how their motion will be affected by disorder.  An extended object has enormously more degrees of freedom in its motion than a simple point particle, and it is not obvious that disorder can lock the entire object in place and prevent it from thermalizing with the rest of the system.

In this work, we will show that a system of extended objects can indeed fail to thermalize, in a natural generalization of many-body localization.  We will describe a concrete mechanism by which disorder can immobilize extended objects, resulting in a localized spectrum.  The strategy will be one of dimensional reduction, in which we phrase the motion of extended objects in terms of the propagation of lower-dimensional objects.  For example, we can frame the motion of a string in terms of the propagation of point-like kink defects.  When the disorder is sufficiently strong, these effective point particles can exhibit conventional many-body localization.  As we will argue, the localization of such internal modes will preclude any overall coherent motion of the string, due to the inability of the string to maintain causal contact between its different segments.  In this way, standard many-body localization of point particles can lead to the localization of a string, locking it in place along some one-dimensional locus.  (We note a phenomenological connection with the recently discovered ``fractonic lines," which are locked in place along one-dimensional paths due to a set of conservation laws, as opposed to disorder-driven localization \cite{pai}.)  This is true regardless of the dimensionality of the space in which the string can move. More formally, a convergent locator expansion can be constructed around each possible location of a string.  Similarly, we can phrase the motion of a two-dimensional surface in terms of the propagation of one-dimensional wavefronts, which in turn can be reduced to the motion of kink defects.  The localization of extended objects is thereby understood in a hierarchical fashion. We introduce a basic phenomenology of many body localized systems with extended excitations, similar to the `$\ell$-bit' description \cite{serbyn, huse} familiar from point particle MBL but supplemented by a local constraint.  We also propose a type of out-of-time-order string correlator as a diagnostic of such a phase.  All our arguments are `asymptotic' ($i.e.$ valid up to possible rare region corrections), but they generalize the study of MBL to a host of settings beyond those which have hitherto been considered. 

Our analysis will provide insights into a number of different situations, including domain walls in ferromagnets, loop excitations in three-dimensional topological phases, and flux lines in three-dimensional type-II superconductors.  We will discuss the implications which are unique to each situation, along with the caveats that are necessary for application of our analysis.  We emphasize in particular that applying the concept of localization protected order \cite{order} to type-II superconductors implies that flux line localization could stabilize superconductivity to temperatures (more precisely, energy densities) where in thermal equilibrium the system would be in the normal state, and could thus offer an entirely new route to high temperature type-II superconductivity.  The phenomenology of such a phase has a family resemblance to flux pinning by defects in vortex glass phases \cite{fisher,ffhuse}.  However, we emphasize that flux pinning is classical and thermodynamic in origin, pinning the objects only to the minima of the potential landscape.  An applied external field can then trigger depinning, leading to characteristic signatures in non-linear conductivity \cite{ffhuse}. In contrast, the extended object localization described here prevents the system from thermalizing at all.  The absence of motion is driven by quantum effects, rather than by free energy considerations, and the extended objects need not be locked in place at local minima of the potential. Rather, objects cannot move because moving would take the system `off energy shell,' and there is no heat bath in the problem to supply or absorb energy.  A superconductor with localized flux lines will therefore have sharp differences in phenomenology from a vortex glass phase.  For example, an applied external field will not lead to depinning, leading to very different behavior in the nonlinear conductivity.

\section{Localization of Strings}

\subsection{Reduction to Point Particle Localization}

\begin{figure*}[t!]
 \begin{minipage}[b]{0.45\linewidth}
 \centering
 \includegraphics[scale=0.55]{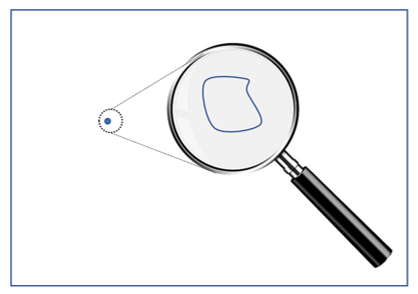}
 \caption{For a string much smaller than the system size, we can zoom out to a coarse-grained description in which the string effectively looks like a point particle, in which case the conventional theory of localization applies.}
 \label{fig:small}
 \end{minipage}
 \hspace{1cm}
 \begin{minipage}[b]{0.45\linewidth}
 \centering
 \includegraphics[scale=0.55]{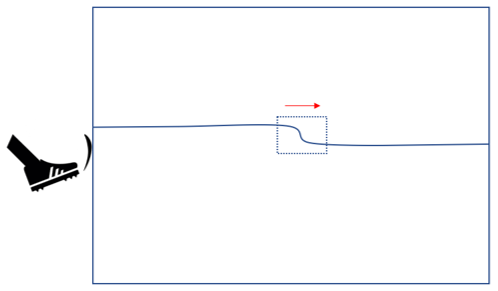}
 \caption{For a long string stretching across the system, the two ends are not in direct causal contact.  If the left end is given a kick, the right end will remain unaffected unless a kink propagates across the entire system.}
 \label{fig:large}
 \end{minipage}
 \end{figure*}

We begin by studying the motion of a single string in a disordered medium, in the absence of any inter-string interactions.  This can be thought of as the study of `Anderson' localization of strings.  The extension to interacting strings will be discussed later. Implicit in all the following analysis is the idea that the string in question has a non-zero line tension, $i.e.$ large changes in the length of the string take the system `off energy shell' and are therefore suppressed. Before focusing on any specific physical system, we can argue on simple physical grounds that the problem of string localization can generically be reduced to an analysis of point particles moving in $(1+1)$ dimensions.  This portion of our analysis is completely general and pictorial.  We will make the discussion more precise by introducing a specific model in a subsequent section. Our discussion is implicitly formulated for a string that lives on a {\it lattice}. The motion of a string through {\it continuous} space maps to a many body localization problem in the $(1+1)$-dimensional continuum, which has subtleties associated with the unbounded nature of the local Hilbert space \cite{rahul, continuum} that we do not discuss here.  We also assume that the string moves according to a local Hamiltonian, as will be the case in the specific model studied in the next section.

For a generic system with string-like excitations on a lattice, there are two important limiting cases to consider: small closed loops, and macroscopic lines threading through the entire system.  These two limits will illustrate the two basic types of processes which can move a generic string governed by a local Hamiltonian.  For a small closed loop, with dimensions much smaller than the system size, we can simply zoom out to a coarse-grained description.  At long distances, we can neglect any internal structure and treat the entire loop as a point particle, as in Figure \ref{fig:small}.  We are then free to focus only on the center of mass motion of the entire string.  Disorder in the system will result in a random potential landscape for this effective point particle, and sufficiently strong disorder will result in localization.  More generally, if the size of the string is smaller than the size of local operators in the Hamiltonian, then the center of mass of the string can be regarded as an effective point particle moving according to a local Hamiltonian, which can be localized by disorder via ordinary point particle localization.  Since the overall size of the string is small, we can then conclude that the entire string is exponentially localized to within some bounded region.

This sort of analysis will of course break down when the string under consideration is bigger than the size of local operators in the Hamiltonian, such as in the case of a string stretching all the way across the system.  For such large strings, their extended nature clearly cannot be ignored.  We must therefore adjust our strategy.  Instead of zooming out to a point particle description, we will use the large size of the string to our advantage.  The key idea is that motion of a long string through local operations generically requires the propagation of point-like kinks down its length.  For example, consider a string threading through the entire system, as pictured in Figure \ref{fig:large}.  The two ends of this string are not in direct causal contact.  If one end of the string moves, through either some internal fluctuation of the system or an external force, the other end will not immediately know to move in unison.  It will take a finite amount of time for any signal to propagate down the string and set the opposite end in motion.  This signal will take the form of a pulse, or kink, which in a clean system will propagate away from the site of the initial kick at a speed determined by the tension of the string.  In this way, information can be exchanged between different parts of the string, and the string as a whole will move.
 
This mechanism for the motion of a string depends crucially on the ability of signals to propagate down its length, in the form of kink defects.  In a translationally invariant system, the propagation of these effective quasiparticles will proceed unhindered.  In a disordered system, however, the kinks will see a random effective potential, which can obstruct their mobility.  When the disorder is sufficiently strong, the kinks will exhibit many-body localization and will fail to propagate.  Indeed, we are faced with an effective one-dimensional problem of kinks moving along a string, so the effects of disorder will be particularly strong.  (Note that, while disorder always localizes non-interacting particles in one dimension, an interacting one-dimensional system can also exhibit a delocalized phase, so localization will require the disorder to be larger than some critical value).  When the kinks become localized, it is no longer possible for signals to propagate down the string.  The different parts of the string will have no means of communicating and will be unable to establish an overall coherent motion.  Rather, when the string is given a kick at some spatial location, it will result only in a localized oscillation of that small segment of string.  There is no local operation which can result in overall motion of the string.  Note that this argument applies not only to strings of macroscopic extent, but to any string larger than the size of local operators in the Hamiltonian.  In this case, any local operation can only effect a change on a particular segment of the string, and overall string motion will necessarily require propagation of information down its length, by considerations of causality.

The above argument rules out any mode of motion in which the string `wiggles' as it moves through the system, with each segment pulling on its neighbors.  One may then worry about strictly uniform translations of the string, in which all segments move in the same direction at precisely the same speed, so that no information needs to propagate along the string.  However, we can rule out such motion on physical grounds.  First of all, setting up such motion would require a fine-tuned nonlocal operator which provides identical impulses to each segment of the string.  This type of motion certainly cannot be set up by generic local fluctuations.  And even if such motion were artificially engineered, it would not be stable.  Interaction with a disordered landscape will quickly cause different segments of the string to move at different velocities.  In other words, strictly uniform motion of the string will not be an eigenstate of the system. Indeed, a uniform translation of the string will typically take the system `off energy shell' by an amount of order $\sqrt{L}$, where $L$ is the length of the string, based on the central limit theorem (assuming short-range correlated background disorder).  All string motion will thus necessary involve wiggling of the string, $i.e.$ the propagation of internal modes.  By the argument of the previous paragraph, such motion will not be possible when the kinks moving along the string become localized.  We can therefore conclude that, at sufficiently strong disorder, a string will be localized, in the sense of being restricted to exist within a small distance of a particular one-dimensional locus. In particular, note that not only will the `center of mass' of the string stay near its original position, but the {\it orientation} of the string will also remain close to its original orientation, since rotations of the string (while maintaining approximately constant length) also require motion of kinks along the direction of the string, and such motion is prohibited when the kinks are localized. 

\begin{figure}[t!]
 \centering
 \includegraphics[scale=0.55]{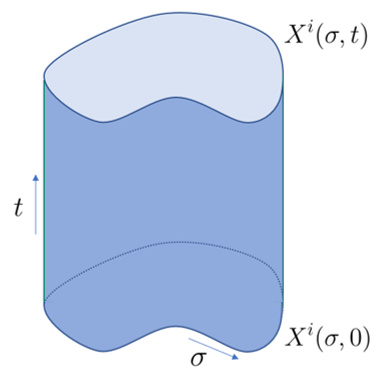}
 \caption{The motion of a single string can be described in terms of fields $X^i(\sigma,t)$, representing the coordinates of each segment, defined on the $(1+1)$-dimensional worldsheet of the string.}
 \label{fig:worldsheet}
 \end{figure}

Our arguments above have relied on a mapping of the problem of string localization onto a one-dimensional many-body localization problem.  One way to formalize this concept is to make use of a `worldsheet' representation of the string, as commonly used in string theory, which explicitly describes the motion of a string in terms of a $(1+1)$-dimensional field theory \cite{polchinski}.  We begin by parametrizing the string by some internal coordinate $\sigma$.  We then label the spatial coordinate of a segment at parameter $\sigma$ and time $t$ as $X^i(\sigma,t)$, as in Figure \ref{fig:worldsheet}.  The Lagrangian for the string will then be a functional of these spatial coordinates, such that the action takes the form:
\begin{equation}
S = \int dtd\sigma \,\mathcal{L}[X^i(\sigma,t)]
\end{equation}
For the study of many-body localization, however, working with actions and their associated path integrals is troublesome, due to a localized system's ability to remember initial conditions at arbitrarily long times.  It is in general more convenient to work with a Hamiltonian, which we can write in the form:
\begin{equation}
H = \int d\sigma \,\mathcal{H}[X^i(\sigma),P^i(\sigma)]
\end{equation}
where $P^i(\sigma)$ is the canonical conjugate variable to $X^i(\sigma)$.  In either formulation, the problem of string propagation is mapped onto a $(1+1)$-dimensional field theory.  For a string propagating in $d$-dimensional space, we can simply regard the coordinates $X^i$ as $d$ fields defined on the $(1+1)$-dimensional worldsheet, parametrized by $\sigma$ and $t$.  The problem is then reduced to an ordinary field theory analysis, where the quantized particles of the fields $X^i$ correspond to modes propagating along the string.  For strong enough disorder, these particles will be localized \cite{imbrie} with respect to the internal coordinate $\sigma$.  In other words, there will be no signals propagating down the string. This formalizes the notion that, in the presence of strong disorder, different segments of the string will not have the causal contact necessary for coherent motion, so the string as a whole will be localized.

We note that, up to now, we have only discussed the localization of a single isolated string, which proceeds via the many-body localization of its point-like kink defects.  However, it seems plausible that this analysis also extends to a system of multiple strings with short-range interactions.  In general, introducing weak coupling between two localized systems will leave the combined system in a localized phase.  Intuitively, if a system cannot serve as a heat bath to thermalize itself, it also cannot serve as a heat bath for another system.  Similarly, if a string cannot reach internal thermal equilibrium of its own kinks, it cannot serve as a heat bath which thermalizes the kinks of another string.  We therefore expect that the localization of strings will be robust against the introduction of weak inter-string interactions.  Explicitly demonstrating this stability to inter-string interactions would require a more complicated analysis, which we leave to future work.  But under this plausible assumption, we can conclude that higher-dimensional systems with a non-zero density of strings and weak short range interactions between strings should then also admit an MBL phase, related to the localized phase of a single string via a type of `wire construction,' at least up to the potential rare region issues discussed in \cite{stability}.  We note that the issue of rare regions is not any more severe in the case of extended objects than in conventional point-particle localization.  While the objects being localized are now extended, the rare regions have no such extended structure.  The dominant mechanism of instability of localization will be the growth of small point-like thermal volumes, just as in the case of point particle localization.   Whether such rare region effects can  be brought under control and a rigorous proof of MBL established, either for true randomness or for quasiperiodicity, is an interesting question that is beyond the scope of the present work.

\subsection{Example: Localized Strings in the Three-Dimensional Toric Code}
\label{3dtc}

We now illustrate the ideas discussed in the preceding section by specializing to a particular model with line-like excitations: the toric code ($Z_2$ lattice gauge theory) in three spatial dimensions. We work on the cubic lattice, with Pauli spins on each link.  In the stabilizer code limit of the theory, we can write the toric code Hamiltonian as:
\begin{equation}
H_0 = -\sum_+ A_+\prod \sigma^x - \sum_\square B_\square\prod\sigma^z
\label{stab}
\end{equation}
where we have allowed for disordered coefficients $A$ and $B$, which we assume are random positive values.  The first sum is over all vertices of the lattice, with each term a product of the six $\sigma^x$ operators touching the vertex.  The second sum is over all plaquettes, with each term a product of the four $\sigma^z$ operators going around the plaquette.

For simplicity, we take our system to have trivial topology.  There is then a unique ground state specified by $\prod_+ \sigma^x = \prod_\square \sigma^z = 1$ for all vertices and plaquettes of the lattice.  More generally, any eigenstate can be labeled by a bit-string representing the values of $\prod_+ \sigma^x$ and $\prod_\square \sigma^z$.  A state with $\prod_+ \sigma^x = -1$ on site $i$ is regarded as having a particle excitation living on that site.  These particles must always occur in pairs, since the product of all vertex operators in the system must be $1$.  Similarly, a state with $\prod_\square \sigma^z = -1$ on a plaquette can be regarded as having a string excitation passing through the center of that plaquette.  These string excitations are forced to form closed loops, constrained by the fact that the product of plaquette operators around a small cube must always be $1$.

The stabilizer code Hamiltonian of Equation \ref{stab} features static particle and string excitations, without any dynamics.  As such, the eigenstates of this Hamiltonian are trivially localized.  The important question to ask is whether this localization is robust to the introduction of perturbations.  Let us consider a perturbed Hamiltonian of the form:
\begin{equation}
H = H_0 + h\sum_{-}\sigma^z + g\sum_{-}\sigma^x + \cdot\cdot\cdot
\label{pert}
\end{equation}
where the two added sums are over individual spins on the links of the system.  All other perturbations are comparatively less important.  The $h$ term flips the value of $\prod_+ \sigma^x$ on two adjacent sites.  As such, this term acts as a hopping operator for the point particles of the theory.  Note that the Hamiltonian of Equation \ref{pert} is explicitly a sum of terms describing the dynamics of point particles and strings.  The $A_+$ and $h$ terms represent the rest and kinetic energy of point particles, while the $B_\square$ and $g$ terms represent the line tension and kinetic energy of strings, respectively.  In this sense, the point particles and strings are almost decoupled in this Hamiltonian.  The only interaction arises from the fact that the point particle number operator is constructed from the same $\sigma_x$ operators as the string kinetic term (and similarly for the string measurement operator and particle kinetic term).  This interplay simply results in mutual statistics between the two sectors of the theory.  This statistical interaction can be simply taken into account in the wavefunction picture by adding appropriate phase factors.  One can write a wavefunction which almost separately describes configurations of both point particles and strings, with the exception of phase factors determined by their relative positions.

By standard localization arguments, the point particles will remain localized provided $h$ is much smaller than the strength of the disorder in $A_+$.  More formally, one could determine this by studying wavefunctions in the $\sigma_x$ basis, in which states are labeled by point particle positions and strings are represented simply by phase factors.  One can then construct a locator expansion in this basis, which will converge at sufficiently strong disorder (neglecting rare region effects).  Since the point particles can be localized in straightforward fashion, we will ignore this sector entirely from here on, focusing only on the string excitations.  The portion of the Hamiltonian describing the dynamics of the string sector is:
\begin{equation}
H_{str} = -\sum_\square B_\square\prod\sigma^z + g\sum_- \sigma^x
\label{stringham}
\end{equation}
where the first term is the static string Hamiltonian of the stabilizer code, and the second term is a small perturbation.  The $g$ term flips the value of $\prod_\square \sigma^z$ on the four plaquettes touching a link, thereby creating a closed loop around that link.  Equivalently, this term can be regarded as a kinetic term for a string, deforming it from one path to an adjacent one.

\begin{figure}[t!]
 \centering
 \includegraphics[scale=0.5]{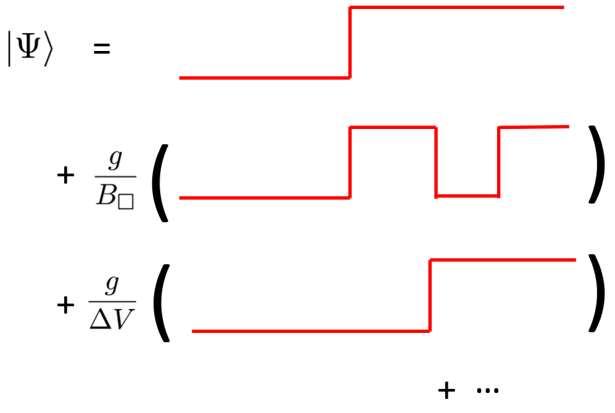}
 \caption{The eigenstates of the perturbed Hamiltonian will be a mix of states in which new kinks have been created and old kinks have moved.  For sufficiently strong disorder, both types of processes will involve significant energy change, $\Delta E\gg g$, so that the locator expansion converges.}
 \label{fig:wavefunction}
 \end{figure}
 
We can now find the eigenstates of the new Hamiltonian perturbatively.  Starting from the stabilizer code eigenstates with a single fixed string configuration, the $g\sigma^x$ perturbation will mix in slightly deformed configurations, as indicated schematically in Figure \ref{fig:wavefunction}.  Such deformations will be weighted down by powers of $g/\Delta E$, where $\Delta E$ is the change in energy associated with deforming the string.  There are two main types of deformation processes.  The first type of deformation involves the creation of two new kink defects in the system ($i.e.$ a particle-antiparticle pair) or the deletion of two existing kinks.  Because the string generically has a finite line tension, there will be a finite energy cost to creating these particles, set by the typical value of $B_\square$.  As long as $g$ remains much smaller than the typical $B_\square$, creation of particle-antiparticle pairs will be an `off-resonant' virtual process, with exponentially small weight in the wavefunction.  To a first approximation, these types of processes can be disregarded entirely.  In principle one could worry about accidental cancellations between the tension and disorder terms, leading to near zero denominators. However, it is important to note that string line tension is controlled by the typical value of $B_\square$, whereas the disorder strength is controlled by the standard deviation in $B_\square$, which we henceforth denote $\Delta V$.  By making these two scales very different, such as by taking the `taut string' limit where $\Delta V \ll \langle B_\square \rangle$, we can essentially eliminate such accidental cancellations.

The second, more important type of deformation process depicted in Figure \ref{fig:wavefunction} corresponds to the motion of an existing kink along the string.  Such processes will be weighted down by factors of $g/\Delta V$, since $\Delta V$ is the typical change in potential seen by the moving defect.  We have now reduced the problem to a one-dimensional point particle localization analysis, as expected.  By the usual logic of conventional MBL, sufficiently strong disorder (sufficiently small $g/\Delta V$) will cause the kinks to be localized at their position along the string.  (We note that a similar kink localization problem has recently been studied explicitly in the context of topological insulator edge states \cite{yangzhi}.)  Shifted kink states will have weights in the wavefunction which are exponentially small in the distance $\Delta r$ of a kink away from its original position.  As a whole, weights in the wavefunction will be exponentially small in the sum of the displacements of all kinks, $i.e.$ proportional to $\exp(-\frac{1}{\xi}\sum|\Delta r|)$, where $\xi$ is the kink localization length.  We now see that two simple physical assumptions, namely localization of kinks and a finite line tension, imply that the perturbation $g$ will not deform the string wavefunction significantly from its original static configuration.  Each eigenstate of the system is smoothly connected to a state with a fixed location for the string.  In other words, the locator expansion of the eigenstates is convergent.

\begin{figure}[t!]
 \centering
 \includegraphics[scale=0.5]{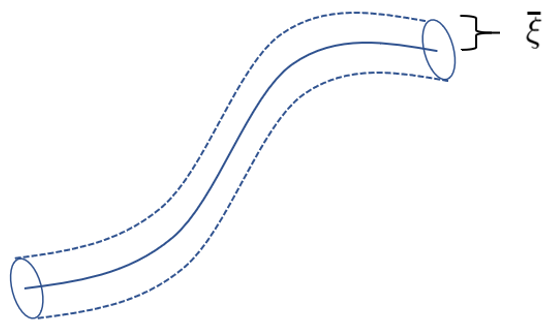}
 \caption{Almost all of the weight of the wavefunction, up to exponential tails, will correspond to string configurations inside a tube of width $\overline{\xi}$ around the original string locus.}
 \label{fig:tube}
 \end{figure}

In particular, it is worth noting that the wavefunction of the string will have negligible weight outside a narrow tube around the original one-dimensional locus, as depicted in Figure \ref{fig:tube}.  Extending the string transversely by a distance $d$ would require one of two processes.  Either the length of the string grows by $2d$ through a kink creation process, which is weighted down by a factor of $(g/\langle B_\square\rangle)^{2d}\equiv e^{-d/\xi_B}$, or $d$ existing kinks move past a particular point on the string, which is weighted down by a typical factor of $(g/\Delta V)^{d/n} \equiv e^{-d/\xi_V}$, where $n$ is the average density of kinks along the string.  The wavefunction will therefore be negligible outside a tube of width $\overline{\xi} = \textrm{max}\{\xi_B,\xi_V\}$.

\subsection{Localization of a Single String}

In order to make the preceding analysis more rigorous, we can construct several mappings from the problem of string localization onto established results in the literature, which will serve to confirm our arguments that strong disorder can result in the localization of a single extended string. Note that (crucially) a localization problem for a single string maps onto a {\it many body} localization problem for pointlike objects.

We first construct an explicit mapping from various types of string excitations onto one-dimensional models of point like excitations.  We begin with the two-dimensional Ising model, which has the simplest such mapping, the logic of which we will carry over to other models, like the three-dimensional toric code.  The Hamiltonian of the disordered ferromagnetic Ising model takes the form:
\begin{equation}
H = -\sum_{\langle ij\rangle} J_{ij}\sigma^z_i\sigma^z_j - \lambda \sum_i \sigma^x_i
\end{equation}
where $J_{ij}$ takes random positive values and we assume $\lambda \ll \langle J \rangle$. We emphasize that the transverse field Ising model in two spatial dimensions is already an interacting system, since unlike it's one dimensional counterpart, it cannot be Jordan-Wigner transformed into free fermions.

\begin{figure}[t!]
 \centering
 \includegraphics[scale=0.3]{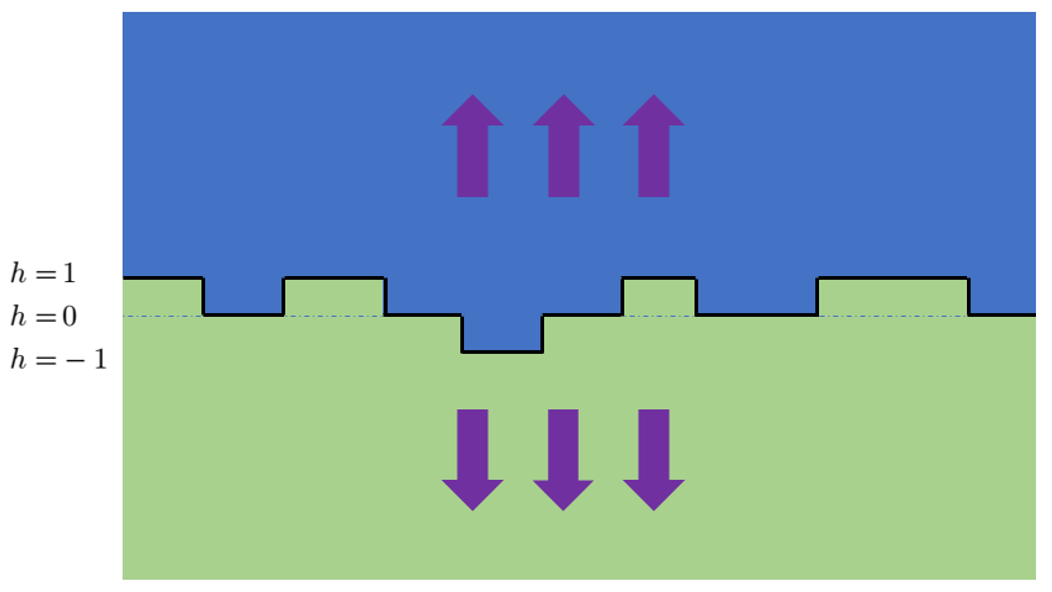}
 \caption{When an Ising domain wall weakly fluctuates around a straight-line configuration, the resulting string can be labeled in terms of a one-dimensional height model.}
 \label{fig:height}
 \end{figure}

To study string localization in the simplest possible setting, we now focus on a state with a single large domain wall running across the entire system in a straight line, dividing the system into halves of opposite spin.  In the limit where the string tension is much greater than the disorder strength, this corresponds to a local minimum of the energy.  We then consider the effective Hamiltonian for fluctuations of the domain wall around this minimum.  The lowest energy excitations correspond to the creation of kink-antikink pairs.  Assuming that these kinks are not too dense, we can label the resulting string configuration in terms of a one-dimensional height model, representing the displacement of the domain wall from the initial straight-line configuration, as depicted in Figure \ref{fig:height}.  Taking the initial configuration to lie in the $x$ direction, the effective Hamiltonian for the displacement takes the form:
\begin{align}
\begin{split}
H = \sum_x\bigg(2J_x(x,h(x)) + 2\overline{J}_y(x,h(x))|h(x+1)-h(x)|\\
-\lambda (\pi^\dagger(x,h(x)) + \pi(x,h(x)))   \bigg)
\end{split}
\end{align}
where $h(x)$ is the height of the domain wall at position $x$, and $\pi^\dagger(x)$ ($\pi(x)$) is the raising (lowering) operator which increases (decreases) the value of $h(x)$.  These operators obey the commutation relations:
\begin{equation}
[h(x),\pi^\dagger(y)] = \pi^\dagger(y)\delta(x-y)
\end{equation}
where we have set Planck's constant to unity.  The symbol $J_x$ denotes the Ising coupling along $x$-directed links (arising from the flat sections of the domain wall), and $\overline{J}_y$ is the Ising coupling on the $y$ links averaged over the height jump at each kink.  The Hamiltonian now explicitly has the form of a one-dimensional system.  As long as we are in the low-energy sector, where the kinks are sparse enough that we can assume the string does not double back on itself, this height representation and its mapping onto a one-dimensional problem is exact.

We can now go one step further, by working in the limit of large line tension, where kink defects are very energetically costly and therefore dilute.  In this limit, where kinks are well isolated, the height will change by at most one unit from one site to the next, such that $\Delta h (x) \equiv h(x+1)-h(x) = 0,\pm 1$.  We can then choose to work with this height \emph{difference}, $\Delta h(x)$, as our fundamental variable.  (Since $h(x)$ was naturally defined on the links of the original lattice, $\Delta h(x)$ is naturally defined on the vertices.)  The local Hilbert space now becomes equivalent to that of a spin-1 system, which importantly is bounded in size.  We denote the effective spin operators of this Hilbert space via tildes, $\tilde{S}_x$, $\tilde{S}_z$, and so on, to avoid confusion with the operators of the original spins of the two-dimensional Ising model.  In this language, we can represent the height difference $\Delta h$ as an $\tilde{S}_z$ operator.  The operator $\pi^\dagger$ raises $h$ on a single link, which raises and lowers $\Delta h$ on two adjacent vertices:
\begin{equation}
\pi^\dagger (x) = \tilde{S}_+(x)\tilde{S}_-(x+1)
\end{equation}
We will also assume for simplicity that $J_x(x,h(x))$ has typical value much larger than its variance, corresponding to the equivalent assumption on line tension.  In this case, $J_x(x,h(x))$ will not vary too much between $h = 0,\pm 1$, so we will take it to be a function of $x$ alone.  The Hamiltonian can then be written as:
\begin{align}
\begin{split}
H = \sum_x\bigg( 2J_x(x) + 2\overline{J}_y\tilde{S}_z(x)& \\
- \lambda(\tilde{S}_x(x)\tilde{S}_x(x+1) +& \tilde{S}_y(x)\tilde{S}_y(x+1))\bigg)
\end{split}
\end{align}
which has the form of a one-dimensional XY model with a random magnetic field, which is known to localize. This is the `non-interacting' limit for kinks, and corresponds to the limit of infinite line tension. At finite line tension kinks can interact via intermediate virtual states that have a height change of more than one unit per vertex. Integrating out such virtual states will generate an effective $\tilde S^z_x \tilde S^z_{x+1}$ interaction. The motion of a single domain wall can thus be mapped onto a one dimensional $XXZ$ spin chain with disorder which, following the rigorous results of Imbrie\cite{imbrie}, should be localized. The obstructions \cite{stability} to extending the arguments of Imbrie to higher dimensions do not apply here, since a state with a single line like excitation is effectively in the zero temperature Hilbert space, and the arguments of \cite{stability} only apply to states at non-zero temperatures. Thus, in the limit of large line tension, the localization of a single string can be rigorously established. 

We note that, as the string tension becomes weaker and kink defects become easier to create, the projection into the $h = 0,\pm 1$ subspace will no longer be sufficient.  A weaker string tension will necessitate keeping higher and higher values of $h$, until eventually we must treat $h$ as a true integer-valued variable, corresponding to the Hilbert space of a quantum rotor, as opposed to a finite-$S$ spin.  Systems with unbounded local Hilbert spaces are outside the scope of rigorous proofs of many-body localization and tend to have obstructions to localization physics \cite{imbrie}.  This illustrates the importance of the assumption of a large line tension, which keeps the low-energy Hilbert space locally finite, while a weak line tension will likely lead to delocalization of kinks and thereby delocalize the entire string.

As another independent confirmation of string localization physics, we can rely on previously established results on strong-disorder renormalization group (RG) in the two-dimensional Ising model, hosting string-like domain wall excitations.  For sufficiently strong disorder, the Ising model is known to flow towards a stable infinite randomness fixed point \cite{motrunich}.  The usual strong-disorder RG prescription, however, focuses on the properties of the ground state, by progressively eliminating the strongest couplings in the system to take us to the low-energy sector.  In contrast, we want to examine the RG flow of parameters in the presence of a domain wall running across the system.  There are certain procedures, such as RSRG-X \cite{rsrgx}, which are designed specifically to analyze the behavior of excited states.  However, the behavior of a state with a single large domain wall can be studied in much simpler fashion.

In order to access the physics of a domain wall state, instead of the true ground state, we can simply choose boundary conditions for our Ising model which force a domain wall into the system.  For example, we can stipulate that all spins are pointed up at $y=\infty$, while all spins are pointed down at $y=-\infty$, in a sort of ``twisted" boundary condition.  These boundary conditions require that there must be at least one domain wall as one moves across the system in the $y$ direction.  Lowering the energy scale by performing strong-disorder RG, and assuming that the typical line tension is much larger than the disorder strength, we will eventually reach a state with a single straight domain wall dividing the system in half.  In the case of untwisted boundary conditions, the couplings flow to infinite randomness \cite{motrunich}.  By locality, specifying boundary conditions at infinity cannot affect the flow of couplings in the bulk, so even in the presence of twisted boundary conditions, the couplings still flow to infinite randomness.  (This type of insensitivity to boundary conditions has been observed previously in certain random spin chains \cite{randomsing}.)  We can then conclude that a state with a single domain wall has dynamics governed by the infinite randomness fixed point.  The resulting state must then be localized. This thus provides an alternative proof of localization of the string-like domain wall in the two dimensional transverse field Ising model. 

Finally, we note that {\it experimentally}, localized domain walls have already been observed \cite{exp4}, consistent with our analysis.

A similar analysis can be extended to strings in the three-dimensional toric code.  In this case, the $B_\square$ term of the Hamiltonian in Equation \ref{stringham} acts as the line tension, giving an energy cost to each length of the string.  Once again, we imagine starting from a straight-line configuration of the string, which corresponds to a minimum of the energy so long as the typical value of $B_\square$ is much larger than the disorder strength.  We then seek to describe fluctuations of the string around this straight line in terms of its kink defects.  At low energies, we can assume a low density of these kinks.  We can then label the deformed string configuration in similar spirit to the previously used height representation.  In the present case, however, there are two independent directions perpendicular to the string in which it can deform.  We must therefore label each point along the string by a two-dimensional vector $\vec{h}$.  Taking the unperturbed string to lie in the $x$ direction, we can write an effective Hamiltonian for the fluctuating string as:
\begin{align}
\begin{split}
&H = \sum_x\bigg(B_{\square,x}(x,\vec{h}(x)) + \overline{\vec{B}}_{\square,\perp}\cdot (\vec{h}(x+1)-\vec{h}(x)) \\
+& g (\pi_y(x,\vec{h}(x)) + \pi^\dagger_y(x,\vec{h}(x)) + \pi_z(x,\vec{h}(x)) + \pi_z^\dagger(x,\vec{h}(x)))\bigg)
\end{split}
\end{align}
where $B_{\square,x}$ represents couplings on plaquettes with $x$-directed normal.  Similarly, $\overline{\vec{B}}_{\square,\perp}$ represents the coupling on perpendicular links, averaged over the change in $\vec{h}$ at each point along the string.  The variables $\pi_y$ and $\pi_z$ represent the lowering operators which decrease the value of $h_y$ and $h_z$, and similarly for the raising operators $\pi_y^\dagger$ and $\pi_z^\dagger$.  Once again, the dynamics of a long string has been mapped explicitly onto a one-dimensional problem.

To map onto the rigorous results of Imbrie \cite{imbrie}, we must once again truncate to a locally finite Hilbert space.  To this end, we assume that the fluctuations around straight-line configuration are weak enough that we can restrict to $\Delta h_y(x) = 0,\pm 1$, and similarly for $\Delta h_z(x)$, with the additional constraint that on any given site, $\Delta h_y$ and $\Delta h_x$ cannot both be non-zero.  We denote the effective spin operators for this Hilbert space as $\tilde{S}_z^{(y)}$, and so on, where the superscript $(y)$ denotes the corresponding direction of fluctuations of $\vec{h}$.  Following along with all steps applied to the two-dimensional Ising model, the str ing Hamiltonian can be written as:
\begin{align}
\begin{split}
\sum_x\bigg( B_{\square,x}(x) + 2\overline{B}_{\square,y}\tilde{S}_z^{(y)}(x) + 2\overline{B}_{\square,z}\tilde{S}_z^{(z)}(x) \\
+ g(\tilde{S}_x^{(y)}(x)\tilde{S}_x^{(y)}(x+1) + \tilde{S}_y^{(y)}(x)\tilde{S}_y^{(y)}(x+1) \\
\tilde{S}_x^{(z)}(x)\tilde{S}_x^{(z)}(x+1) + \tilde{S}_y^{(z)}(x)\tilde{S}_y^{(z)}(x+1)) \\ + \lambda (\tilde{S}_z^{(z)}(x) \tilde{S}_z^{(y)}(x))^2\bigg)
\end{split}
\end{align}
where $\lambda$ is a large positive number and the term proportional to $\lambda$ implements (softly) the constraint that string cannot fluctuate on both directions on any given site. Integrating out virtual states with higher energies will again generate $zz$ interactions between nearest neighbors (including interactions that couple the two spin `species'). Nevertheless, the model maps onto a pair of coupled spin chains with short range interactions and disorder, and without any non-abelian symmetry \cite{symmetry}. We then expect that this problem should many body localize, as problems of this sort generically do, at least for strong enough disorder. 

In addition to the mapping to one dimension, there is another rigorous way to establish localization of the three-dimensional toric code.  The loop-free sector of the toric code can be related by a Wegner duality\cite{wegner1,wegner2} to a three-dimensional transverse field Ising model:
\begin{equation}
H = -\sum_{\langle ij\rangle} J_{ij}\sigma_i^z\sigma_j^z - \lambda\sum_i \sigma_i^x
\end{equation}
The duality with the toric code can be seen by examining the wavefunction in the basis of $\sigma^z$ eigenstates and considering the domain walls between regions of different spin orientations.  In the large $\lambda$ limit, the wavefunction is an equal-weight superposition of all domain wall configurations, $i.e.$ all closed surfaces.  This maps exactly onto the ground state wavefunction of the stabilizer code limit of the toric code (Equation \ref{stab}) in the $\sigma_z$ basis, which is also an equal-weight superposition of closed surfaces.  We therefore see that the deconfined phase of the toric code maps onto the large-field limit of the transverse field Ising model.  Similarly, the strongly confined limit (taking $g\rightarrow\infty$ in Equation \ref{pert}) has no surfaces present in the wavefunction whatsoever, which corresponds to the zero-field limit of the Ising model, where no domain walls are present in the ground state.

Since the loop-free sector maps exactly onto the transverse field Ising model, we can rely on established strong disorder renormalization group results for the latter model\cite{motrunich} to conclude that, for sufficiently strong disorder, the ground state of the three-dimensional toric code is governed by an infinite-randomness fixed point.  We also want to reach the same conclusion for a sector with a single string threading through the system.  As in our discussion of the two-dimensional Ising model, we can fix ourselves to be in this sector through an appropriate choice of boundary conditions, stipulating a single unit of string flux at two opposite ends of the system.  Since this choice of boundary conditions will not affect the flow of local bulk couplings under strong-disorder renormalization group transformations, the system will still flow to infinite randomness, which indicates that the single-string state is localized.

This concludes our section on explicit demonstrations of localization of a single string. Note in particular that the problem of localization of a single extended string maps to a many body localization problem for point particles. We then argue that when many strings are present, then we are coupling together one dimensional MBL systems, and this should still yield a localized state for weak enough couplings, at least up to rare region effects which are beyond the scope of the present work.

\subsection{Phenomenology of States with Localized Strings}

One characteristic feature of a system of particles with full many-body localization is the existence of local conserved quantities which can be used to label all eigenstates of the theory \cite{huse,serbyn,growth}.  For example, consider a one-dimensional chain of Pauli spins with a random uniaxial magnetic field $h_i$, plus a small $xx$ coupling $J$ as well as a $zz$ coupling $\lambda$ (necessary to break integrability). 
\begin{equation}
H = \sum_i (h_i\sigma^z_i + J\sigma^x_i\sigma^x_{i+1} + \lambda \sigma^z_i \sigma^z_i)
\label{chain}
\end{equation}
In the limit $J = 0$, we can label the entire spectrum by the eigenvalues of all $\sigma^z_i$, writing eigenstates as $|\Psi\rangle = |\{\sigma_i\}\rangle$, where $\{\sigma_i\}$ is a string of $\pm 1$ representing the state of each spin.  By specifying one bit at each site, we can specify the entire state.  In this way, the information of the quantum state is encoded in a localized way.  Since such a bit-string is an eigenstate, the value of any bit will not change over time, corresponding to a conserved quantity of the system, $i.e.$ an integral of motion.

In the presence of a nonzero $J$, however, these eigenstates will mix with each other, generating entanglement between the different spins.  The key idea is that, for $J$ small compared to the variance of the disorder, $h_i$, there will still be a set of localized `$\ell$-bits' $\tau_i$ which are conserved quantities.  Furthermore, if the entire spectrum is localized, there will be just as many $\ell$-bits as physical bits (spins), one for each site of the lattice, so these bits can be used to label quantum states.  For weak disorder, the $\ell$-bits will be locally related to the physical bits, such that $\tau_i$ can be expanded as a sum of $\sigma$ operators with weight decaying exponentially away from site $i$.  More formally, since these $\ell$-bits are conserved, they must commute with the Hamiltonian.  We must therefore be able to rewrite the Hamiltonian of Equation \ref{chain} as:
\begin{equation}
H = \sum_i\tilde{h}_i\tau^z_i + \sum_{ij} \tilde{J}_{ij}\tau^z_i\tau^z_j + \sum_{ijk}K_{ijk}\tau^z_i\tau^z_j\tau^z_k + \cdot\cdot\cdot \label{eq:lbitHamiltonian}
\end{equation}
where the coefficients $\tilde{J}_{ij}$, $K_{ijk}$,... fall off exponentially in the separation between sites.  Note that the Hamiltonian only contains $\tau^z_i$, not $\tau^x_i$ or $\tau^y_i$.  This form guarantees that all $\tau^z_i$ commute with $H$ and that specifying the eigenvalue of each $\tau^z_i$ is enough to specify any eigenstate of the Hamiltonian.  As the transition to a thermal phase is approached, the localization length associated with the $\ell$-bits diverges, such that their information spreads out over the entire chain, preventing the system from retaining local memories of its initial conditions.

We now wish to generalize this picture to the case of string localization.  In a phase exhibiting localization of strings (such as the toric code case discussed above), the entire spectrum is in one-to-one correspondence with the set of one-dimensional loci which a string can occupy.  Furthermore, these loci are preserved in time, analogous to the local integrals of motion of localized point particles.  The difference in the present case is that there is no longer a single bit of information preserved at a single location in space, but rather a whole string's worth of information preserved along a one-dimensional path.  We can capture this notion with the same $\ell$-bit Hamiltonian of Equation \ref{eq:lbitHamiltonian}, supplemented by a local constraint $\prod\tau^z = 1$ over each cube of the lattice. (For a discussion of $\ell$-bits in systems with local constraints, see \cite{ChandranBurnell}).  This combination of an $\ell$-bit Hamiltonian and a local constraint gives us localized eigenstates while simultaneously also enforcing that the Hilbert space contains only closed strings.  We can then equivalently say that the system exhibits a set of conserved string operators, $W_C\equiv \prod_C\tau^z$, representing the product of $\ell$-bits along a closed path $C$.  These operators obey $[W_C,H] = 0$ and are in one-to-one correspondence with the possible one-dimensional loci of strings.  The resulting eigenstates of the system are then in direct correspondence with the positions of localized strings, capturing the essential features of the string-localized phase.

\section{Extension to Other Extended Objects}

\begin{figure}[t!]
 \centering
 \includegraphics[scale=0.7]{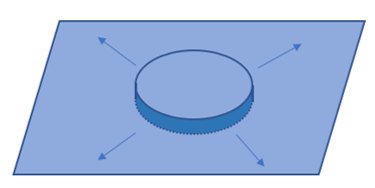}
 \caption{When a two-dimensional surface is given a kick at some location, a signal will propagate outwards as a one-dimensional wavefront.  When all such one-dimensional kinks are localized, the two-dimensional surface itself will be localized.}
 \label{fig:surface}
 \end{figure}

In the previous section, we have argued that a one-dimensional string with finite line tension can be localized simply through the many-body localization of point-like defects moving along that string.  We can make similar arguments for the localization of higher-dimensional extended objects, in a hierarchical manner.  For example, let us study the case of closed two-dimensional surfaces.  When the surface is much smaller than the system size, we can coarse-grain the system until the surface looks like a point particle, and all of the usual localization arguments will carry through (assuming a non-zero surface tension, such that large changes of surface area take the system far off energy shell).  The other case of interest is when we have a very large surface extending across the whole system, as in Figure \ref{fig:surface}.  Once again, if a local kick is given to one section of the surface, the rest of the surface will not immediately know.  Rather, a signal will propagate outwards from the epicenter of the disturbance as a one-dimensional wavefront of kinks.  As in our discussion of strings, if all internal modes on the surface become localized, then no signals can be exchanged between the different sections.  A local kick given to the surface can create a localized oscillation, but it cannot induce overall motion of the surface.  Without communication between the different sections of the surface, coherent motion will be impossible, and the surface will be localized, restricted to within a small distance of its original two-dimensional locus.

In this way, the localization of extended objects proceeds in a hierarchical manner.  A two-dimensional surface moves via the propagation of one-dimensional objects, which in turn rely on the propagation of point-like defects down their length.  Localization of point-like defects is thus sufficient for localization of one dimensional objects, which in turn is then sufficient for localization of two dimensional surfaces. For the (purely academic) analysis of localization of even higher-dimensional objects, a similar hierarchy will hold.  Motion of a $p$-dimensional membrane will require propagation of a $(p-1)$-dimensional wavefront, which in turn can be reduced to a $(p-2)$-dimensional problem and so on until we arrive at a point particle problem.  In this way, the traditional treatment of localization of point particles can be bootstrapped to demonstrate that disorder is capable of localizing any higher-dimensional object (at least up to rare region considerations \cite{stability}).  More formally, one could rewrite the problem in worldvolume language.  We first parametrize the $p$-dimensional object by $p$ internal coordinates $\sigma = (\sigma_1,...,\sigma_p)$.  We then write the spatial coordinate of each patch of the object as $X^i(\sigma,t)$, in close analogy with the analysis of strings.  The Hamiltonian can then be written as:
\begin{equation}
H = \int d^p\sigma\,\mathcal{H}[X^i(\sigma),P^i(\sigma)]
\end{equation}
where $P^i(\sigma)$ is the canonical conjugate to $X^i(\sigma)$.  Once again, we can simply regard the coordinates $X^i$ as $d$ fields defined on the $(p+1)$-dimensional worldsheet.  We then have a normal field theory describing point-like defects of the extended object, and standard localization arguments can be applied.  When disorder is sufficiently strong, the fields $X^i$ will be localized with respect to $\sigma$, and no signals can propagate along the extended object.  Coherent motion will then be impossible, and we can conclude that the extended object is localized.  Following the treatment of the previous section, it will then be possible to create a generalized locator expansion, with conserved bits of information along an entire higher-dimensional locus.

\section{Diagnostics and Characterizations}

Having established general arguments for the existence of a localized phase of extended objects, we now investigate some of the properties which could be used to diagnose such a system.  We first describe some general characterizations which apply to all localized phases of extended objects.  We will then discuss some specific physical situations to which our analysis applies, along with the consequences of localization in each.

\subsection{Out-of-Time-Order String Correlators}

As with any phase of matter, the first line of investigation into extended object localization should be to examine the behavior of correlation functions.  Since we are interested in a non-equilibrium system, it is natural to deploy a powerful recent addition to the toolbox of non-equilibrium physics: the out-of-time-order correlation function (OTOC) \cite{maldacena,channel,roberts,yingfei,yongliang,otocmbl1,otocmbl2,debanjan}. These correlators have been a useful theoretical tool across the non-equilibrium spectrum, from diagnosing the onset of scrambling in a chaotic system, to signaling the breakdown of thermalization in a many-body localized phase.  Furthermore, they are in principle experimentally accessible \cite{measure1,measure7,measure2,measure3}.

The out-of-time-order correlation function between two operators $V$ and $W$ is defined as:
\begin{equation}
F(t) = \langle W^\dagger(t)V^\dagger(0)W(t)V(0)\rangle
\end{equation}
where $V(t) = e^{iHt}Ve^{-iHt}$ is a time-evolved operator in the Heisenberg picture.  The expectation value is taken in some initial state of the system (which is not necessarily an eigenstate, but may be $e.g.$ a product state).  For a thermalizing phase, this OTOC will eventually decay to zero for all choices of $V$ and $W$, with the rate of decay diagnosing chaos and the onset of scrambling \cite{otocmbl2}.  In a many-body localized phase, on the other hand, some of these OTOCs can asymptote to a non-zero constant.

Consider the case of conventional many-body localization, characterized by a set of localized $\ell$-bits $\{\tau_i\}$ specifying the entire spectrum.  If we choose the operator $W$ to be a single $\ell$-bit operator, $W = \tau^z_i$, then we will have $W(t) = W(0)$, since the $\ell$-bits are conserved under time evolution, by definition.  We then choose $V$ to be a local operator far away from site $i$, such that $[W(0),V(0)] = [W(t),V(0)] = 0$.  In this case, the OTOC reduces to:
\begin{equation}
F(t) = \langle V^\dagger V\rangle
\end{equation}
which is simply a constant, with no time evolution.  More generally, we should allow $W$ to be a local operator acting on the physical spins, $e.g.$ $\sigma_i^z$.  We can expand such a local operator in terms of the $\ell$-bit operators:
\begin{equation}
\sigma_i^z = a\tau_i^z + b\tau_i^x + c \tau_i^y + \sum_{j\neq i} a_{ij}\tau_i^z + \cdot\cdot\cdot
\label{expansion}
\end{equation}
where the `off-site' coefficients decay exponentially away from site $i$.  The $\tau_i^{x,y}$ operators are not conserved under time evolution and will have vanishing contribution to the OTOC in the long-time limit.  But for a generic local operator, with nontrivial overlap with $\tau_i^z$, the OTOC will asymptotically reach a nonzero constant, determined by the $a$ coefficients, and decay to this asymptotic value will be power law in time \cite{otocmbl2}.

For the case of a string-localized phase, we no longer simply have $\ell$-bits corresponding to the localized positions of particles.  Rather, we have conserved $\ell$-bit-strings, $W_C = \prod_C\tau^z$, corresponding to the one-dimensional paths of localized strings, as discussed earlier.  If we choose the $W$ operator of the OTOC as a $\tau$-string operator, then clearly at long times we recover $F(t) = \langle V^\dagger V\rangle$, just as in conventional MBL.  Physically, however, we must allow $W$ to be a string number operator in terms of the physical $\sigma$ bits.  Importantly, in the models we consider, such as the toric code and two-dimensional ferromagnets, strings are created at the boundary of \emph{membrane} operators.  We then choose $W = \prod_M\sigma^z$ over a membrane M with boundary C.

Such a physical string number operator can be expanded in terms of the $\tau$-string operators, analogous to Equation \ref{expansion} for point particles:
\begin{equation}
\prod_M\sigma^z = a\prod_C\tau^z + \sum_{C'}a_{CC'}(\prod_C\tau^z)+ (\tau^{x,y}\,\textrm{terms})\cdot\cdot\cdot
\end{equation}
The largest weight in this expansion will be the `$a$' coefficient, corresponding to the $\tau$-string along the path $C$, with all other coefficients decaying exponentially as a function of the `distance' between the paths $C$ and $C'$.  (This can be defined, for example, as the size of the minimal surface spanning the two curves.)  It is then tempting to conclude that the we will see a clear asymptotic constant in the OTOC, dominated by the $a$ coefficient:
\begin{equation}
F(t\rightarrow\infty) = |a|^2\langle V^\dagger V\rangle + \cdot\cdot\cdot
\end{equation}
However, it is important to estimate the size of $a$.  Even when the string is localized to a one-dimensional path, it is possible for each segment of the string to fluctuate away from that path.  Each segment of the string will lead to a depletion of $a$ away from unity, say by a factor of $(1-\epsilon)$.  In the limit of a long string of length $L$, with many fluctuating segments, $a$ will be depleted by a factor of $(1-\epsilon)^L$, which tends to zero as $L$ becomes very large.

In order to counteract these fluctuations, one useful prescription is to ``fatten" our string operators, as has been done in previous studies of string correlators in MBL systems \cite{fat}, by averaging over all string configurations within a one-dimensional tube of width $w$.  This amounts to a coarse-graining of our original system.  We now recall the discussion of our string-locator expansion, where we determined that the wavefunction was negligible outside a tube of size $\overline{\xi}$, as depicted in Figure \ref{fig:tube}.  If we fatten the our string operators beyond the scale $\overline{\xi}$, then the depletion per unit length will be very small, $\epsilon\ll 1$.  By choosing a width $w\gg \overline{\xi}$, the reduction factor per length of the string will be:
\begin{equation}
(1-\epsilon) = (1-e^{-w/\overline{\xi}})\approx 1
\end{equation}
Of course, we must still account for the large size of macroscopic strings, which leads to a much larger reduction factor for the OTOC of the total string:
\begin{equation}
F(t\rightarrow\infty)\sim(1-e^{-w/\overline{\xi}})^L\langle V^\dagger V\rangle
\end{equation}
If we choose $w$ larger than $\overline{\xi}\log L$, then the asymptotic OTOC will remain of order unity.  This is a fairly substantial fattening of the string.  Nevertheless, this logarithmic enhancement of the width is still negligible compared with the length of the string, leaving the path of the string well-defined.  With this choice of `fattened' string operator, we therefore have that the out-of-time-order correlators of this string-localized phase will reach a nonzero asymptotic constant.

It is also possible to use string OTOCs to distinguish a string-localized phase from a conventional MBL system, at least in certain regimes.  Recall that string creation operators are fundamentally \emph{membrane} objects in terms of the underlying physical $\sigma$ spins.  For a string of length $L$, the corresponding operator is a membrane with area of order $L^2$.  In a string-localized phase, the OTOC of this membrane operator is suppressed by a factor $(1-\epsilon)^L$, capturing the fluctuation of the string created at the boundary.  The expectation value is not, however, sensitive to the size of the bulk membrane.  This result is consistent with previous studies on membrane operators in topological phases\cite{gregor}, where it is found that expectation values of membrane operators are insensitive to fluctuations of small loop excitations across the bulk, since small loops are trivial objects which can decay directly into the vacuum.  In contrast, for an MBL phase of conserved point particles, fluctuations of particles across the bulk of the membrane lead to a suppression factor of $(1-\epsilon)^{L^2}$.  When strings are the only excitations in the Hilbert space, such as in a two-dimensional ferromagnet, we can then distinguish between a string-localized phase and conventional point-particle localization by examining the decay of $F(\infty)$ as a function of $L$.  In systems which also have conserved point particles, such as the three-dimensional toric code, these OTOCs will once again decay as $(1-\epsilon)^{L^2}$ at large $L$, due to particle fluctuations across the surface.  However, if the point particles have a much higher gap (or much shorter localization length) than the string excitations, then we expect an intermediate regime of string lengths for which $(1-\epsilon)^L$ decay is observed.

\subsection{Entanglement Behavior}

In terms of quantum entanglement properties, many of the distinguishing features of conventional many-body localization will transfer directly over to the case of extended object localization.  For example, consider entanglement entropy.  An eigenstate of a thermalizing system at finite energy density obeys the ETH, meaning that its entanglement entropy should have thermal behavior, scaling as the volume of a region, $S\sim \ell^d$.  In contrast, a many-body localized system only has entanglement between nearby degrees of freedom, even at finite-energy density, so that the entanglement entropy of a partition depends only on the surface area of the partition, $S\sim \ell^{d-1}$.  This logic, first applied in the case of point particle localization, carries over unchanged to a system with extended object localization, which will therefore exhibit an area law for entanglement entropy of all eigenstates.  Similarly, normal many-body localized phases exhibit slow (logarithmic) growth of entanglement after a quantum quench, since there are no propagating degrees of freedom to enable the rapid spread of quantum information \cite{moore}.  The phases we have discussed here feature complete localization of strings, including localization of signals propagating down those strings.  As such, systems of localized extended objects also have no means of rapidly transferring quantum information, and we expect the same slow logarithmic growth of entanglement entropy.

\subsection{Other Experimental Signatures}

Out-of-time-order correlation functions and entanglement entropy are useful tools for identifying localized phases in numerics and are in principle also experimentally measurable, but they are not the simplest quantities to access in experiments.  More experimentally useful metrics of localization include echo experiments \cite{interferometry}, autocorrelation functions \cite{ngh}, and the decay of revivals of test spins \cite{revivals}. These diagnostics should carry through essentially unchanged to systems with localized extended objects. Transport coefficients also constitute experimentally accessible diagnostics (albeit available only for systems with a conserved charge). For a system exhibiting localization of extended objects, we can conclude that in linear response there will be no transport of any quantum numbers carried by the extended objects.  For example, in a three-dimensional quantum spin liquid (a type of topological phase), spin can be transported by either the point particles or by extended string excitations.  If the disorder in the system is strong enough to localize both types of excitations, then we can conclude that there is vanishing dc spin transport.  Without the string localization mechanism discussed here, a three-dimensional spin liquid would not have strictly zero spin conductivity.  Vanishing spin transport coefficients within a three-dimensional spin liquid would therefore be a direct signature of string localization. In a type-II superconductor, the localization of flux lines implies the absence of transport of magnetic flux across the system, the implications of which we discuss in the next section. Indeed any system with localization of extended objects should have vanishing dc conductivities of appropriate quantum numbers.  An interesting further question to ask would be how such conductivities behave at finite frequency.  In standard many-body localization, the ac conductivity has a rich structure in linear \cite{gopal} and non-linear \cite{khemani} response. Performing the analogous calculation for a system with extended objects seems quite challenging, and it is unclear whether similar results will be obtained.  We leave this as an open question for future investigation.
\label{sec:characterization}

\subsection{Applications}

\subsubsection{Ferromagnets}

Perhaps the simplest setting in which extended objects occur is in a ferromagnet, where we have domain walls between regions of spins aligned in different directions.  In two dimensions, these take the form of loops, while in three dimensions, they take the form of closed two-dimensional surfaces.  In either case, we can characterize the spectrum in terms of these extended objects, and the concepts discussed in the previous sections can be applied without issue.  We then expect that, in the presence of strong disorder, the domain walls of the ferromagnet will be localized.  Having domain walls locked in place will make it much harder for a ferromagnet to change its magnetization in response to an external magnetic field, which ordinarily proceeds via domain wall motion.  We therefore expect the magnetization of a sample to be largely frozen in, changing only via much slower processes, such as the rearrangement of entire domains.  We note that a single localized domain wall has already been observed experimentally in a two dimensional system \cite{exp4}. 

\subsubsection{Three-Dimensional Topological Phases}

Another setting in which extended objects naturally occur is in three-dimensional topological phases.  In this case, we should bear in mind that the spectrum contains both loop excitations and point particles.  However, we can simply apply a standard point particle localization analysis to argue that, with sufficiently strong disorder, both point particles and loop excitations will exhibit localization.  As argued earlier, weak coupling between two localized systems will not result in delocalization, so we can conclude that the entire spectrum of the topological phase is localized.  Experimentally, we should observe a vanishing thermal conductivity, along with vanishing conductivities for any other quantum numbers carried by the loop and particle excitations of the topological phase.

Another important consequence of localization physics is that it allows highly excited eigenstates of a system to exhibit types of order which are more usually associated with ground states or equilibrium systems, such as symmetry breaking order and topological order \cite{order}.  Furthermore, these eigenstates can exhibit order which is forbidden by theorems of equilibrium statistical mechanics, such as spontaneous symmetry breaking at non-zero energy density in one dimension, or two-dimensional topological order at finite energy density.  In three dimensions, it is known that equilibrium topological order only partially survives at finite temperature, with a fraction of the ground state topological entanglement entropy \cite{chamon}.  In a fully localized system, however, we conjecture that full topological order should survive even at finite energy densities.  Our analysis of extended object localization thereby opens the door to new three-dimensional examples of localization protected order, such as in the three dimensional toric code discussed in Sec. \ref{3dtc}.

\section{Localized Superconductors in Three Dimensions}

We now discuss in detail a particularly important example of a phase with localized string excitations - a localized superconductor in three spatial dimensions. It has already been argued \cite{longrange} that superconductors in a hypothetical two-dimensional world (with strictly two-dimensional gauge field) can exist in a localized phase wherein all excitations are localized and the system has zero thermal conductivity, with vanishing electrical resistance. However, the extension of this analysis to real superconductors in three spatial dimensions is nontrivial, because three-dimensional superconductors admit vortex {\it line} excitations, and (prior to the present manuscript) it was not clear how localization of extended objects should be described. Armed with our results on extended object localization, however, we are now ready to tackle the problem of localization in three-dimensional superconductors.

\subsection{Lattice Gauge Theory for a Superconductor}

We begin by introducing a lattice gauge model serving as an effective theory for a three-dimensional superconductor, which will turn out to be closely related to the toric code model, allowing us to apply our previous results directly.  Our analysis is an adaptation of \cite{HOS} to three spatial dimensions, formulated entirely on a lattice.
 
A superconductor occurs when electrons, interacting through a long-range Coulomb repulsion, form a pair condensate.  Physically, this long-range interaction between electrons is mediated by photons moving at the speed of light.  For typical solid state applications, however, this velocity is so large that we can ignore any retardation effects and regard the electrons as interacting through an instantaneous long-range potential, without any reference to a photon field.  We can then write an effective Hamiltonian within the electronic sector (ignoring phonon coupling for now) as:
\begin{equation}
H = -\sum_{\langle ij\rangle}t_{ij}c^\dagger_ic_j + e^2 \sum_{i\neq j} \frac{c^\dagger_ic_i c^\dagger_j c_j}{|r_i-r_j|} - \mu\sum_i c^\dagger_i c_i
\label{ham}
\end{equation}
where $i,j$ are site labels of the underlying lattice, and $c^\dagger$ ($c$) is an electron creation (annihilation) operator.  The first term is the kinetic energy of a non-interacting system, capturing the hopping of electrons from site to site.  (Only nearest neighbor hopping is considered for simplicity.)  The second term is the instantaneous Coulomb interaction.  We have also included a chemical potential term, working in the grand canonical ensemble.  By starting from this effective Hamiltonian in our localization analysis, instead of electrons coupled to a photon, we thereby avoid the issues associated with localization in the continuum \cite{continuum}.

The elimination of the photon field comes at a cost, however.  Whereas the interaction of electrons with the photon field can be written in purely local terms, the effective Hamiltonian of Equation \ref{ham} features explicit long-range interactions.  Many of the standard treatments of many-body localization no longer apply in the presence of long-range interactions, though recent work has shown that localization can survive in such systems, at least in certain cases\cite{longrange}.  Luckily, for our purposes, we can easily rewrite the Hamiltonian in a form which is both local and defined on the lattice.  While the Hamiltonian of Equation \ref{ham} descended from the coupling to the physical continuum photon field, the same effective Hamiltonian would be obtained from coupling to a \emph{lattice} $U(1)$ gauge theory, with speed of light large compared to the characteristic scales in the problem.  To this end, we can write an equivalent Hamiltonian for the electrons as:
\begin{align}
\begin{split}
H = -\sum_{\langle ij\rangle}t_{ij}&c^\dagger_i e^{ia_{ij}}c_j + \sum_i a_{0,i} ((ev)\,c^\dagger_i c_i - \textrm{div}(e)_i) \\
- \mu\sum_i c^\dagger_i &c_i+ \frac{1}{2v^2}\sum_{\langle ij\rangle} (e_{ij})^2 - \sum_\square \cos(\textrm{curl}(a)_{\square})
\label{gaugeham}
\end{split}
\end{align}
where $a_{ij}$ is a compact $U(1)$ vector gauge field defined on the link between site $i$ and site $j$ (recalling that $i$ and $j$ are site labels and not spatial indices) with canonically conjugate $e_{ij}$ representing the lattice electric field.  Meanwhile, the timelike component $a_{0,i}$ is defined on site $i$.  The final term is a sum over all plaquettes of the lattice, with the curl defined as the discrete line integral of $a_{ij}$ around each plaquette.

The parameter $v$ in the Hamiltonian represents the velocity of the lattice photon.  When $v$ is large compared to other velocity scales in the problem, the dynamics of the gauge field become unimportant.  We can then simply resort to an electrostatic analysis, in which we write $e_{ij} = \phi_j - \phi_i$ for potential $\phi_i$ defined on each site.  By integrating out $a_{0,i}$, we also obtain a lattice Gauss's law, $\textrm{div}(e)_i = (ev) c^\dagger_i c_i$, which combines with the potential formulation to give us a lattice Poisson equation, $\nabla^2\phi = -(ev)\,c^\dagger_ic_i$.  Solving this equation will give us the usual electrostatic potential for point particles, falling off as $\phi\sim 1/r$ away from charges.  Using the potential form for $e_{ij}$ in Equation \ref{gaugeham}, we will recover the desired instantaneous Coulomb interaction, plus a term representing the energy of the (now static) magnetic field:
\begin{align}
\begin{split}
H = -\sum_{\langle ij\rangle}&t_{ij}c^\dagger_i c_j + e^2 \sum_{i\neq j} \frac{c^\dagger_ic_i c^\dagger_j c_j}{|r_i-r_j|} \\
&- \mu\sum_i c^\dagger_i c_i - \sum_\square \cos(\textrm{curl}(a)_{\square})
\end{split}
\end{align}
Note that the magnetic field term is defined on the lattice and has been compactified, which is not necessarily a good description of magnetic flux in a metal.  However, in the superconducting phase, which is our primary concern, the magnetic flux will be quantized and tied directly to the lattice vortices of the condensate, so this lattice description of flux will be accurate.  Furthermore, a $2\pi$ flux line in a superconductor is `trivial' except for its associated energy cost ($i.e.$ it has no mutual statistics with any other excitations), so the compact description will be adequate for our purposes.  Note also that whereas in two dimensions the compact theory was unstable to confinement \cite{HOS}, this problem is absent when the theory is formulated in three spatial dimensions. In this way, the lattice gauge theory of Equation \ref{gaugeham} serves as a suitable starting point for the analysis of an electronic superconductor.  This theory has the advantage of being both purely local and defined entirely on a lattice, which will allow for standard arguments of many-body localization to be applied.

We now wish to study the Higgsed lattice model in which the $U(1)$ gauge group has been broken down to $\mathbb{Z}_2$.  Physically, this occurs when the electrons form a Cooper pair condensate, such that $\langle c^\dagger c^\dagger\rangle\neq 0$ and particle number is no longer conserved (though it remains conserved mod 2).  At each site, we no longer have an integer particle number $n_i$, but rather a $\mathbb{Z}_2$-valued particle number variable, which we denote $\tilde{\tau}^z_i$, representing the Bogoliubov quasiparticles.  Acting on such condensed states, the fermionic creation operators $c^\dagger_i$ reduce to acting as $\tilde{\tau}^x_i$ operators.  (Note that, in addition to on-site Pauli commutation relations, we must also impose anti-commutation of $\tilde{\tau}^x$ on different sites, to capture the Fermi statistics.)  Within the condensed phase, the magnetic flux is restricted to $0$ or $\pi$, such that $e^{i\,\textrm{curl}(a)} = \pm 1$.  Up to a choice of global phase, we can then write the remaining gauge degrees of freedom as a $\mathbb{Z}_2$-valued gauge field, $e^{ia_{ij}}\rightarrow \sigma^z_{ij}$.  Making these replacements in Equation \ref{gaugeham}, the resulting effective Hamiltonian in the condensed phase takes the form:
\begin{align}
\begin{split}
H = -\sum_{\langle ij\rangle}t_{ij}\tilde{\tau}^x_i\sigma^z_{ij}&\tilde{\tau}^x_j + \sum_i \sigma_{0,i}\tilde{\tau}^z_i\prod\sigma^x -\mu\sum_i\prod\sigma^x \\
&+g\sum_i\sigma^x_i - \sum_\square \prod\sigma^z
\end{split}
\end{align}
where the products are over spins on all links touching the corresponding site or plaquette, as appropriate to the sum.  The variable $\sigma_{0,i}$ is the residual $\mathbb{Z}_2$ Lagrange multiplier enforcing the $\mathbb{Z}_2$ Gauss's law of the theory, $\tilde{\tau}^z_i\prod\sigma^x = 1$.  The $g$ term is descended from the $(e_{ij})^2$ term of Equation \ref{gaugeham}, which acts to change the gauge variable $e^{ia_{ij}}$, thereby serving as a kinetic term for the loop excitations.  In this way, we obtain a $\mathbb{Z}_2$ lattice gauge theory for a three-dimensional superconductor.  The excitation spectrum includes fermionic point particles (Bogoliubov quasiparticles), captured by the first line of the Hamiltonian, and loop excitations ($\pi$ flux lines), captured by the second line.  The coupling in the first term of the Hamiltonian also insures that the point particles pick up a minus sign upon braiding around a loop, as is appropriate for Bogoliubov particles moving around a $\pi$ flux.  This $\mathbb{Z}_2$ gauge theory is essentially a three-dimensional toric code model, but coupled to fermionic matter made up of pointlike excitations. Since all sectors containing pointlike excitations can exhibit conventional MBL, the only challenge is to demonstrate localization of flux lines. Our earlier analysis of flux line localization in the three-dimensional toric code then carries over largely unchanged to show that a (lattice) superconductor can be in a localized phase where both the quasiparticle and the flux line excitations are localized. 

As a final subtle point, systems which break a continuous symmetry are typically characterized by a large manifold of degenerate states, which would complicate the localization analysis.  In the superconducting case, this `set' of states is parametrized by the phase $\phi$ of the condensate, $\langle c^\dagger c^\dagger\rangle \sim e^{i\phi}$. One may therefore worry about a `zero mode' where the phase of the order parameter globally changes. However, the symmetry in question is a \emph{gauge} symmetry, and all such states are actually physically equivalent, eliminating concerns about such zero modes. (Additionally, of course, global zero modes do not contribute to internal rearrangements of the system, and as such are irrelevant for localization/delocalization). Meanwhile plasmon modes corresponding to finite wavelength phase fluctuations (which are relevant for the discussion of localization) are gapped by the Higgs mechanism, and can be localized as shown in \cite{longrange}.

\subsection{Phenomenology of a Localized Superconductor} 

The `localized superconductor' phase identified above will be characterized by an {\it infinite} dc charge conductivity yet a zero dc thermal conductivity. Note in particular that when a current flows through a superconductor, any flux lines experience a transverse force. If the flux lines are delocalized, then the motion of flux lines gives rise to a finite resistance in the system, so that the superconductivity is no longer perfect. In the presence of sufficiently strong disorder, however, our analysis indicates that the flux lines can exist in a localized phase, which will significantly alter the phenomenology of a type-II superconductor.  In particular, a current will no longer be able to drive flux lines across the system, which is necessary for current dissipation.  A type-II superconductor with localized flux lines should therefore still exhibit perfect superconductivity. At the level of phenomenology, this is similar to the vortex glass phase that occurs in disordered equilibrium superconductors \cite{fisher, ffhuse}. However, unlike the vortex glass the `localized superconductor' phase discussed herein is not in thermodynamic equilibrium, and the vortex lines are not necessarily pinned at minima of the potential. As a result, applied external field will not trigger depinning, and the non-linear conductivity of the system should be very different to the vortex glass. Furthermore, previous studies of localization protected order have shown that localized systems can continue to exhibit order at energy densities which would disorder an equilibrium system.  The localization of flux lines should therefore lead to perfect superconductivity at higher `temperatures' and magnetic fields than would be possible in an equilibrium system.  Such suppression of resistivity in a type-II superconductor could have important technological implications. 

Finally and somewhat provocatively, we note that experimentally there is evidence for long-lived non-equilibrium populations of quasiparticles in real superconductors \cite{qp1, qp2, qp3, qp4, qp5, qp6}, and these present a considerable challenge for engineering devices based on superconducting qubits. A localized superconducting state (which is a non-equilibrium superconducting state that is stabilized by disorder) could provide an intriguing, if exotic, explanation for this experimental mystery. A detailed investigation of this possibility, perhaps using one or more of the diagnostics discussed in the previous section, is left to future work.

We close by discussing several subtleties in the application of our results to real three-dimensional superconductors.  First of all, magnetic flux fundamentally occurs in the continuum, as opposed to the previous examples where everything could be defined on a lattice, and there are well-known difficulties associated with localizing excitations in the continuum.\cite{continuum}  But while magnetic flux generically occurs in a continuum, the quantized flux lines through a superconductor are directly tied to the vortices of the superconducting condensate.  Since the electrons in a solid move on a lattice, the vortices will also only exist at discrete locations in the system, as captured by the effective lattice gauge theory.  To what extent the continuum nature of the physical electromagnetic field presents an obstruction to MBL is thus unclear. As another important caveat, we note the effects of {\it rare regions}, which present their own obstructions to localization in high dimensions and can lead to thermalization at very long times \cite{stability}.  Whether rare region effects can be brought under control (perhaps in models with quasiperiodicity) is beyond the scope of the present work and is left to future investigations.

Additionally, superconductivity is generically a low temperature phenomenon, and we do not expect it to survive to infinite energy densities. Arguments from \cite{deroeck} suggest that such a many-body mobility edge may endow the model with a finite (but long) relaxation time. Whether that is the case, and if so, how long the relaxation time is, are also questions that we defer. We also note that for analytical convenience, we have assumed here that the `flux lines' are narrow on the lattice scale whereas in a real superconductor, a vortex core will typically be larger than the lattice scale. This does not make an important difference to our analysis (one could imagine simply moving to a coarse-grained lattice), but we note it nonetheless as an idealization.  Finally, as a practical matter, we note that the disorder must be strong enough to localize the flux lines while not being so strong that it destroys the superconductivity entirely. Localized superconductivity should thus be something of a `goldilocks' phenomenon, requiring disorder that is neither too strong nor too weak.

\section{Conclusions}

In this work, we have shown that a disordered system of extended objects can fail to thermalize, via the localization of its eigenstates.  We have demonstrated that the localization of extended objects can be reduced to a standard point particle localization analysis, by studying the behavior of signals propagating along the extended object.  When all such internal modes are localized, the entire extended object will itself be localized.  This extends the notion of localization to a host of phases of matter where line (and membrane) excitations naturally arise. We have discussed the basic phenomenology of localized systems with extended excitations and have introduced a type of out-of-time-order string correlator as a diagnostic of such a phase.

Our analysis has implications for a variety of disordered systems, including ferromagnets, topological phases, and superconductors. Our work thus opens the door for new types of localization protected order in three dimensions.  In particular, our analysis suggests that three-dimensional superconductors can exist (at least up to rare region subtleties) in a localized phase where flux lines, quasiparticles, and plasmons are all localized. The resulting localized phase will exhibit a divergent dc charge conductivity and zero thermal conductivity, as well as supporting stable non-equilibrium populations of quasiparticles.  Such `localization of vortex lines' could also stabilize superconductivity to higher magnetic fields and temperatures than are possible in equilibrium systems.  A detailed investigation of this possibility, and its potential experimental and technological consequences, is left to future work. 

\section*{Acknowledgments}

The authors acknowledge useful conversations with A. Burin, Sudip Chakravarty, Yang-Zhi Chou, Sarang Gopalakrishnan, David Huse, Han Ma, Abhinav Prem, Leo Radzihovsky, Wojciech De Roeck, Albert Schmitz, and S.L. Sondhi. This material is based in part (R.N.) upon work supported by the Air Force Office of Scientific Research under award
number FA9550-17-1-0183. M. P. is supported partially by NSF Grant 1734006 and partially by a Simons Investigator Award to Leo Radzihovsky.

\end{document}